\documentclass[preprint]{aastex}
\newcommand{\be}{\begin{equation}}
\newcommand{\ee}{\end{equation}}
\newcommand{\bvc}[1]{\mbox{${\bf e}_{#1}$}}         % basis vector
\newcommand{\uvc}[1]{\mbox{\boldmath ${\hat{#1}}$}} % unit vector
\newcommand{\vc}[1]{\mbox{\boldmath ${\bf #1}$}}    % regular vector

\shorttitle{Polarization Sweeps}
\shortauthors{Hibschman and Arons}

\begin{document}

\title{Polarization Sweeps in Rotation Powered Pulsars}
\author{Johann A.~Hibschman \altaffilmark{1,2} and
        Jonathan Arons \altaffilmark{1,2,3}}
\affil{University of California, Berkeley \altaffilmark{4}}
\altaffiltext{1}{Theoretical Astrophysics Center}
\altaffiltext{2}{Physics Department}
\altaffiltext{3}{Astronomy Department}
\altaffiltext{4}{Address correspondence to J. Hibschman, TAC, 601 Campbell, Berkeley 94720-3411.  email: {\tt johann@leporello.berkeley.edu}}
%\affil{Berkeley, CA}
%\email{johann@leporello.berkeley.edu}
%\and
%\author{Jonathan Arons \altaffilmark{1}}
%\affil{UC Berkeley Astronomy Department}
%\altaffiltext{1}{Theoretical Astrophysics Center}
%\email{arons@astroplasma.berkeley.edu}
% UCB Department of Astronomy, Department of Physics, and TAC}

\begin{abstract}
We re-examine the characteristic polarization angle sweep of
rotation-powered pulsars and calculate the expected deviations from
this sweep caused by aberrational effects and by polar-cap current
flow.  We find that in addition to the previously known phase shift of
the entire sweep by $\Delta \Phi = -4 r/R_L$, aberration shifts the
polarization angle itself by $\Delta \Psi = -(10/3) (r/R_L) \cos
\alpha$.  Similarly, current flow above the polar cap shifts the
polarization sweep by $\Delta \Psi = (10/3) (r/R_L) (J/J_{GJ}) \cos
\alpha$, potentially providing a method of directly measuring the
magnitude of the current.  The competition between these two effects
produces a potentially observable signature in the polarization angle
sweep.  Although these effects may appear similar to orthogonal mode
shifts, they are an independent phenomenon with distinct observational
characteristics.
\end{abstract}

\maketitle

\section{Introduction}

The characteristic S-curve polarization sweeps exhibited by pulsar
emission have spawned a large body of research, both theoretical and
observational.  Ever since the highly-successful rotating vector model
of \citet{rc}, hereafter RC, fitting pulsar polarization profiles has
been a standard method of constraining the underlying pulsar geometry.
Some pulsars, however, have polarization behavior which does not fit
the standard S-curve.  The millisecond pulsars, for example, seem to
have noisy and on average flatter polarization sweeps than normal
\citep{xilouris}.  In attempts to understand these deviations, several
relativistic and plasma effects have been proposed which would modify
the sweep through plasma propagation effects
\citep{ba86,mckinnon97}, through aberration of the beaming
direction from strict parallelism \citep{blas}, or through multiple
interacting orthogonal modes \citep{mckinnon}.

Such effects become progressively more important as the radius of
emission (or the polarization-limiting radius) approaches the light
cylinder distance, $R_L \equiv c/\Omega$.  $r/R_L$ is very small in
the polar cap emission model for pulsars ($r \approx R_*$), a model
consistent with most observational constraints on $r$ \citep{rankin}.
Since $R_L$ is significantly smaller for millisecond pulsars, due to
their faster rotation, we expect these effects to be proportionally
larger.

However, most of the departures from the pure RC model previously
considered have depended on effects of second-order or higher in
$r/R_L$, such as the field sweepback perturbation of
\citet{barnard86}.  The only first-order effect which has been
considered is the aberrational shift due to the co-rotational
velocity, examined by \citet{blas}, hereafter BCW.  In this paper, we
re-examine that effect and analyze an additional first-order effect
due to current flow above the polar cap.

This is a phenomenological study, guided by the success of the
rotating vector model.  That model contains three chief assumptions:
that the observed radiation is beamed along the direction of the
magnetic field lines, that the polarization of the radiation is at a
fixed angle to the radius of curvature of the field lines, as is the
case for vacuum curvature radiation, and that the underlying magnetic
field is a dipole.  To go beyond this, we relax some of these
assumptions in physically-motivated ways.  Keeping the assumption that
the polarization is related to the curvature of the field lines, we
include aberrational change in the beaming direction and the effects
of currents on the underlying dipole field.

After taking these effects into account, we find that aberration
delays the phase of the polarization sweep with respect to the center
of emission and, in a plot of polarization angle vs. rotation phase,
shifts the entire sweep downwards, while current flow shifts the
entire sweep upwards.  If both effects are present, they may lead to
sharp jumps in the polarization angle near the edges of
current-carrying zones.

\section{Polarization Perturbations}

Since we assume that the polarization is related to the acceleration of
the emitting particles and that the emission is beamed along the
direction of motion, the particle velocity field is the core
quantity examined in this paper.  Particles follow the magnetic field,
so any perturbations to that field are reflected in the velocity.

We apply the same general method to several perturbations.  Starting
with a simple dipole field, we add a perturbation to find the new
velocity field of the emitting particles.  From that field, we can
calculate the radius-of-curvature vector of the particles at each
point in space.  Then, for any pulsar phase, we find the position
where the velocity field at a given radius $r$ points towards the
observer, evaluate the radius of curvature vector at that point, and
project it onto the sky to find the observed polarization angle.

To obtain analytic expressions, these calculations are done to first
order in $r/R_L$.  This naturally gives the difference $\Delta \psi$
between the normal polarization angle sweep, $\psi(\Phi)$ and the
perturbed sweep, $\psi'(\Phi) = \psi(\Phi) + \Delta \psi(\Phi)$.  More
details of this procedure are given in Appendix \ref{ap:method}.

\begin{figure}
  % \figurenum{fig:vectors}
  \plotone{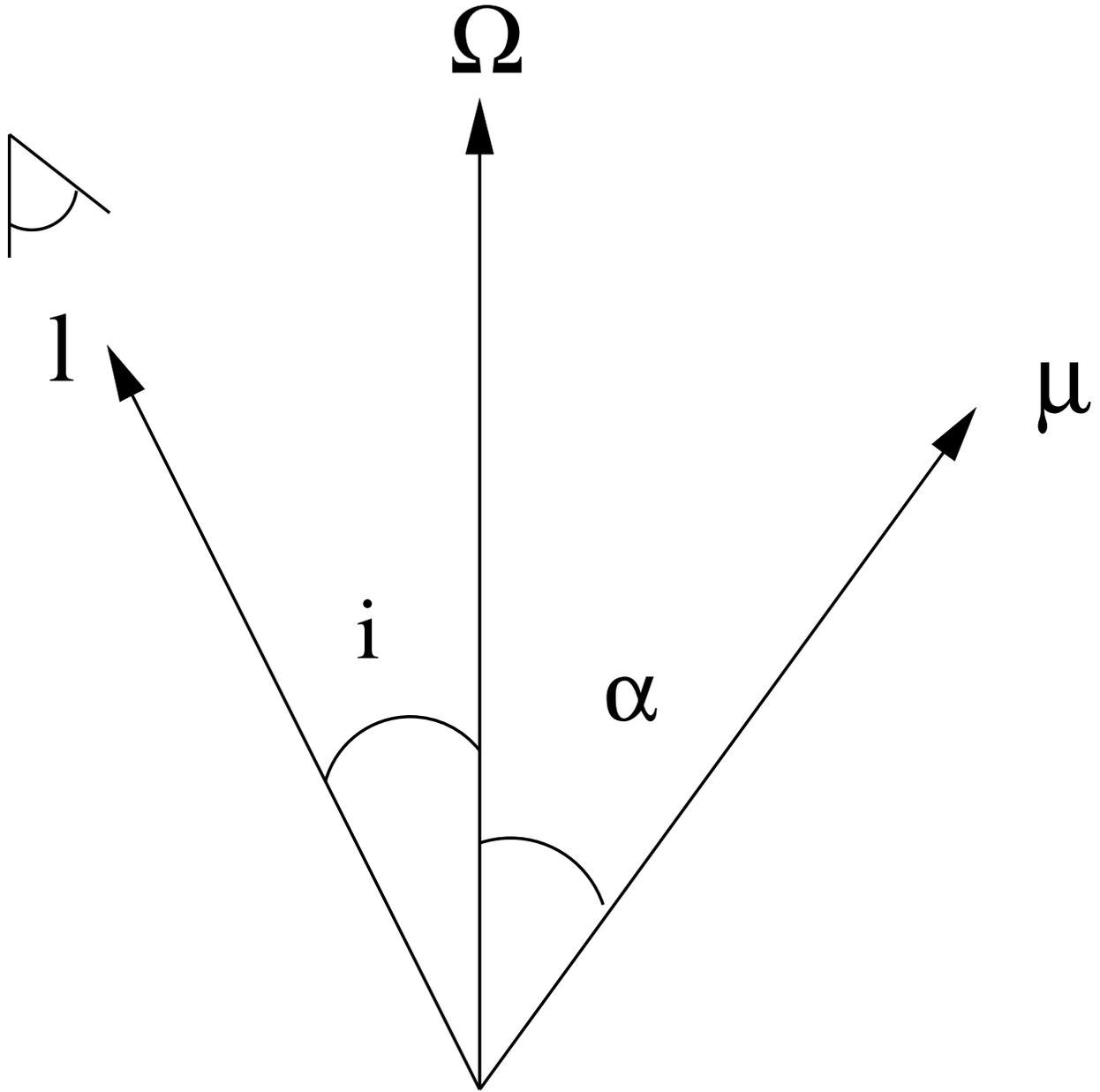}
  \caption{Basic Angles.  $\alpha$ is the angle between the rotation
  axis, $\vc{\Omega}$, and the magnetic moment, $\vc{\mu}$.  $i$ is
  the angle between the observer LOS, $\vc{\ell}$, and the rotation axis.
  \label{fig:vectors}}
\end{figure}

Two angles define the basic geometry of a pulsar: the angle $\alpha$
between the rotation axis ($\vc{\Omega}$) and the magnetic moment
($\vc{\mu}$), and the inclination angle $i$ between the rotation axis
and the line of sight to the observer ($\uvc{\ell}$), see Figure
\ref{fig:vectors}.  The angle of closest approach, the minimum angle
between $\vc{\mu}$ and $\vc{\Omega}$, is then $\beta \equiv i -
\alpha$.

These calculations are done in magnetic-centered coordinates, defined
so that the $z$ axis ($\theta = 0$) is along the magnetic moment
$\vc{\mu}$, the rotation axis $\vc{\Omega}$ is fixed in the $x-z$
($\phi=0$) plane, and the vector to the observer $\uvc{\ell}$ is also in
the $x-z$ plane at $t=0$.

We have also numerically calculated the expected polarization angle
sweeps for each perturbation examined.  Given a modified velocity
field, we directly solved for the new emission point, computed the
radius of curvature vector, and compared it to the rotation axis to
find a polarization angle.  These numerical results confirmed
the analytic perturbation theory.

\subsection{Offset dipoles}

One suggested perturbation to the magnetic field is an offset dipole,
where the center of the dipolar field is offset from the center of the
star, e.g. \citet{chen}.  This has negligible effect on the
polarization sweep.  In an offset dipole, the magnetic origin revolves
about the center of the neutron star with a radius of $\delta < R_*$.
In turn, in the frame of the magnetic field, the observer appears to
move in a circle of identical radius.  The electrodynamics in these
field-centered coordinates is the same as in the star-centered case,
so the only change in the observed emission is that due to the
effective motion of the observer.  This motion and the resultant
changes are tiny, on the order of $\delta/d \ll 1$, where $d$ is the
distance to the observer.

\subsection{Polar Field Aligned Current}

Most models of the pulsar polar cap include a current of charged
particles streaming along the field lines, a current approximately
equal to the Goldreich-Julian density moving at the speed of light.
This current induces a magnetic field of $\vc{B}_1 = (r/R_L)
(J/J_{GJ}) B \cos \alpha \; \uvc{e}_\phi$; this field is purely
toroidal with respect to the magnetic axis.

From Appendix \ref{ap:current}, the resulting polarization angle shift is
\begin{equation}
  \Delta \psi = \frac{10}{3} \frac{r}{R_L} \frac{J}{J_{GJ}} \cos \alpha
    \left( 1 - \frac{7}{40} \sin^2 \theta_0 \right) \: {\rm radians}
  \label{eq:psi1}
\end{equation}
where $\theta_0$ is the magnetic colatitude of the ($0^{th}$-order)
emission point, see Appendix \ref{ap:unperturbed}.  At the center of
the pulse, $\theta_0$ has its minimum value, $\theta_0 \approx (2/3)
|\alpha - i| = (2/3) |\beta|$.  At the boundary of the last closed
field line, $\theta_0$ reaches its maximum, $\theta_0 \approx (2/3)
\sqrt{r/R_L}$.  For $r/R_L$ small, the range of $\theta_0$ is also
small, and the perturbation is mostly flat across the observed pulse.

In terms of the orbital phase, the perturbation is
\begin{equation}
  \Delta \psi \approx \frac{10}{3} \frac{r}{R_L} \frac{J}{J_{GJ}}
    \cos \alpha \left(
    \frac{27}{32} + \frac{5}{32} \cos \beta \right) - \frac{25}{192}
    \frac{r}{R_L} \frac{J}{J_{GJ}} \sin 2 \alpha \sin i \sin^2 \Omega t.
  \label{eq:current-approx}
\end{equation}
To get this form, we used the
approximation formulae of Appendix \ref{ap:approximations}, assumed
that $\Omega t$ was small enough that $\cos \Omega t \approx 1 - (1/2)
\sin^2 \Omega t$, and simplified the constants by taking $28/45 \approx
5/8$, which is good to 0.4\%.  Here, $t=0$ corresponds to the center
of the pulse profile, where the slope of the polarization angle curve
is greatest.

In terms of observables, $r/R_L \approx 1.2^\circ \, P_{0.1}^{-1}
r_{100}$, where $r_{100}$ is the emission radius in units of 100 km
and $P_{0.1}$ is the pulse period in units of 0.1 seconds, so the
magnitude of this shift is $4.0^\circ \, P_{0.1}^{-1} (J/J_{GJ})
r_{100}$.

\begin{figure}
  \plottwo{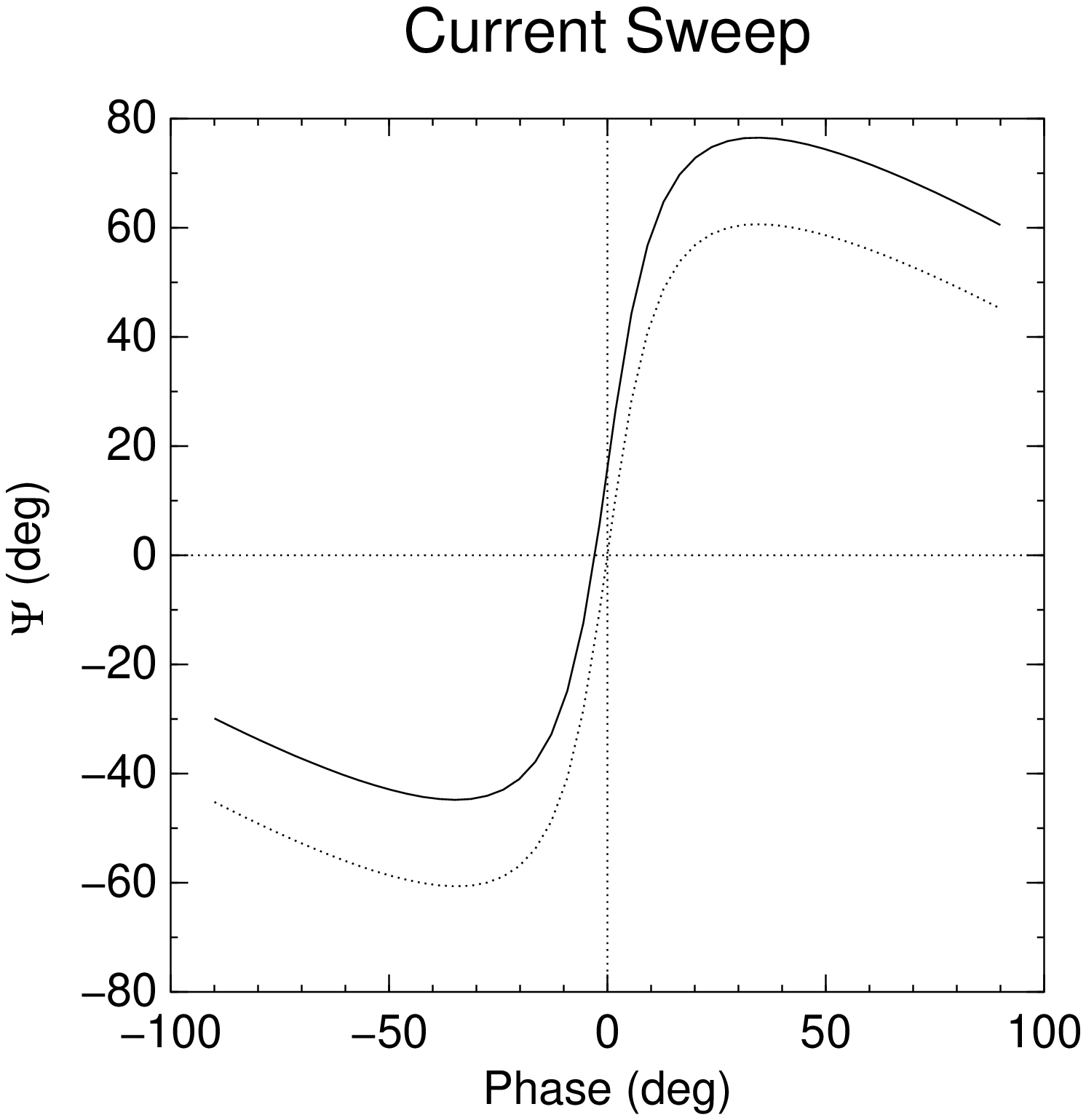}{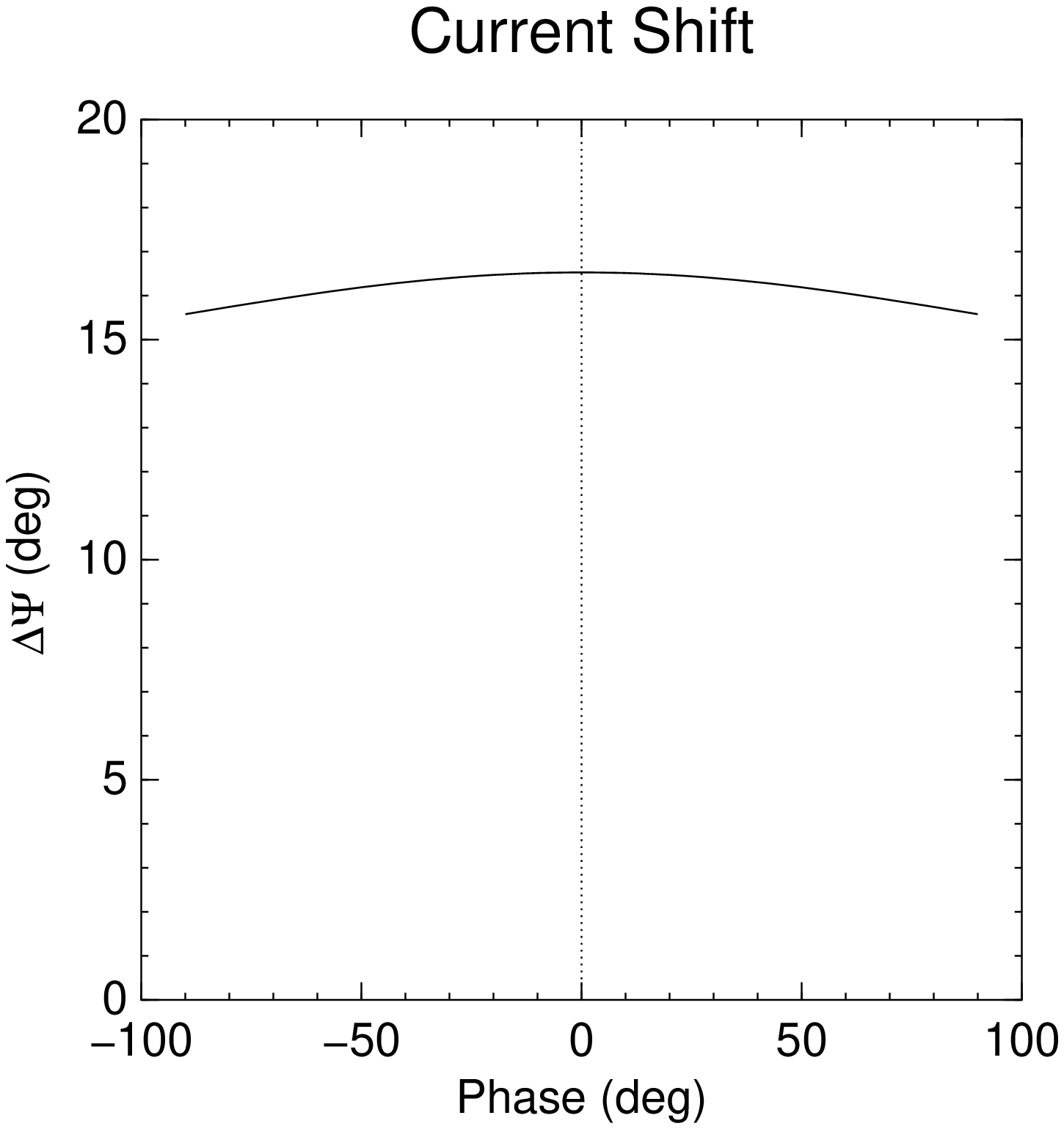}
  \caption{Perturbation to polarization angle, $\Delta\psi$, due to a
  Goldreich-Julian current as a function of longitude.  All angles are
  in degrees.  The dotted line is the unperturbed sweep, while the solid
  line is the perturbed sweep.  $r/R_L = 0.1$, $\alpha = 30^\circ$,
  and $i=35^\circ$.
  % The solid line is the numerical result, the dotted the theoretical.
  \label{fig:current}}
\end{figure}

\subsection{Aberration}

As discussed by \citet{blas}, the rotation of the magnetosphere itself
affects the velocities of particles moving along field lines.  Inside
the light cylinder, the field lines are stationary in the frame
co-rotating with the star.  However, as particles move along these
lines, their paths curve both because of the curvature of the field
line itself and because the frame of those field lines rotates with
the star.  Even if the field lines were straight, the paths of the
particles moving along them would be helices in the inertial frame.
This extra curvature generates a perturbation of the instantaneous
particle directions equal to the co-rotation velocity, $\vc{\Omega}
\times \vc{r}$.

To lowest order, this perturbation shifts the polarization angle by
\begin{equation}
  \Delta \psi = -2 \frac{r}{R_L} \frac{\sin \alpha \cos
    \phi_0}{\sin \theta_0} + O(\frac{r}{R_L}).
\end{equation}
Since $\sin \theta_0$ is generally tiny, on the order of the polar cap
size, $\theta_c \equiv \sqrt{r/R_L}$, this perturbation is of order
$\sqrt{r/R_L}$, a half-order lower than the constant shift produced
by the current.

This shift has the same form as a phase shift of the entire
polarization sweep by
\begin{equation}
 \Delta \Phi = -3 \frac{r}{R_L}
\end{equation}
with respect to the geometric center of the pulse phase, as was found
previously by BCW.  Since aberration advances the emission itself by
$\Delta \Phi = r/R_L$ due to simple beaming, the total phase shift
with respect to the emission is $\Delta \Phi = -4\, r/R_L$.

From numerical investigations, representing this effect as a phase shift
closely matches the perturbation.  The analytic method
presented here only gives the first order correction (in $r/R_L$) to
the sweep; higher-order terms quickly become important, since
aberration adds appreciable curvature to central field lines which
formerly had none.  Representing the change as a phase shift better
reflects these higher-order terms, while giving the same first-order
result.

\begin{figure}
  \plottwo{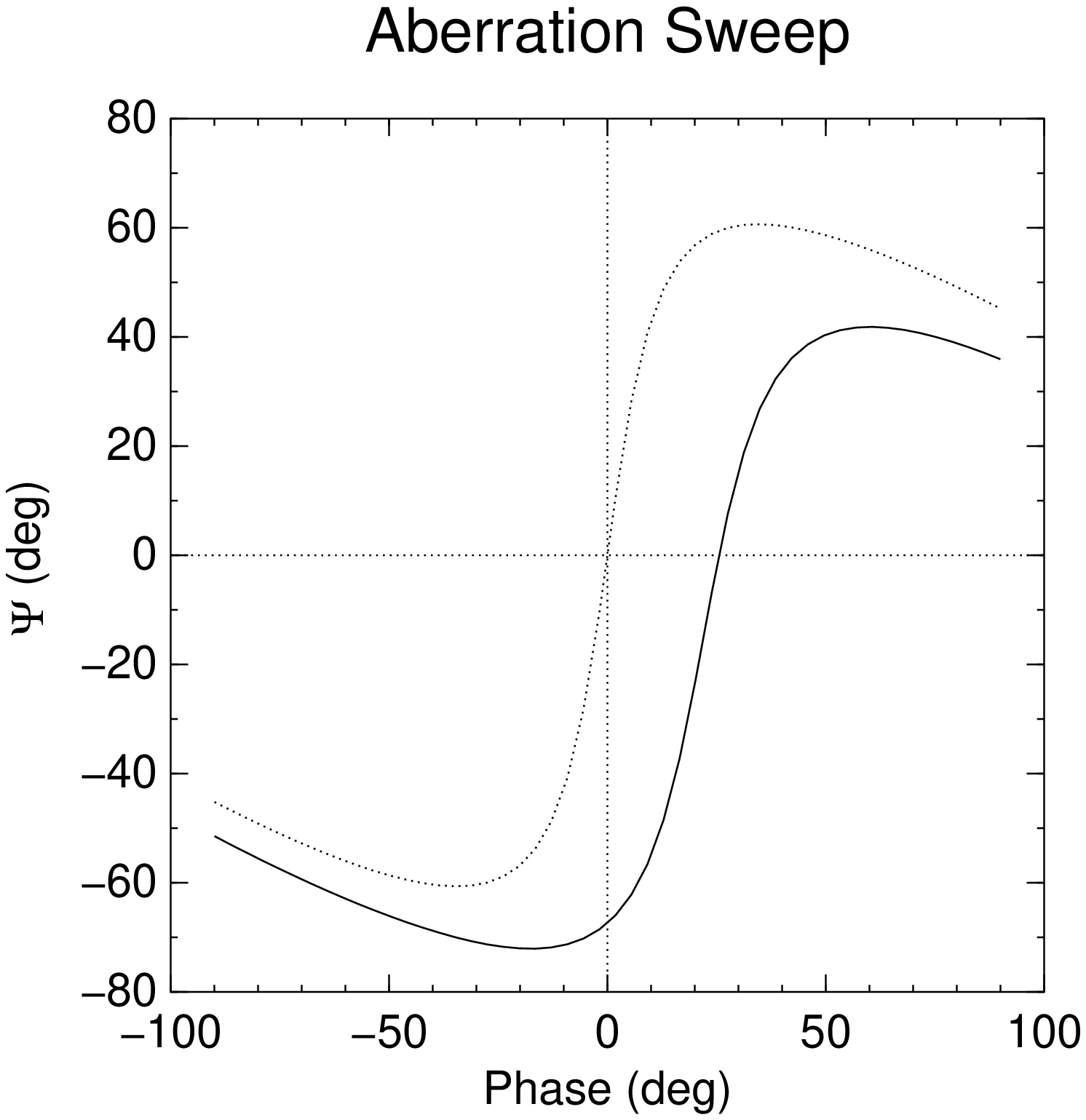}{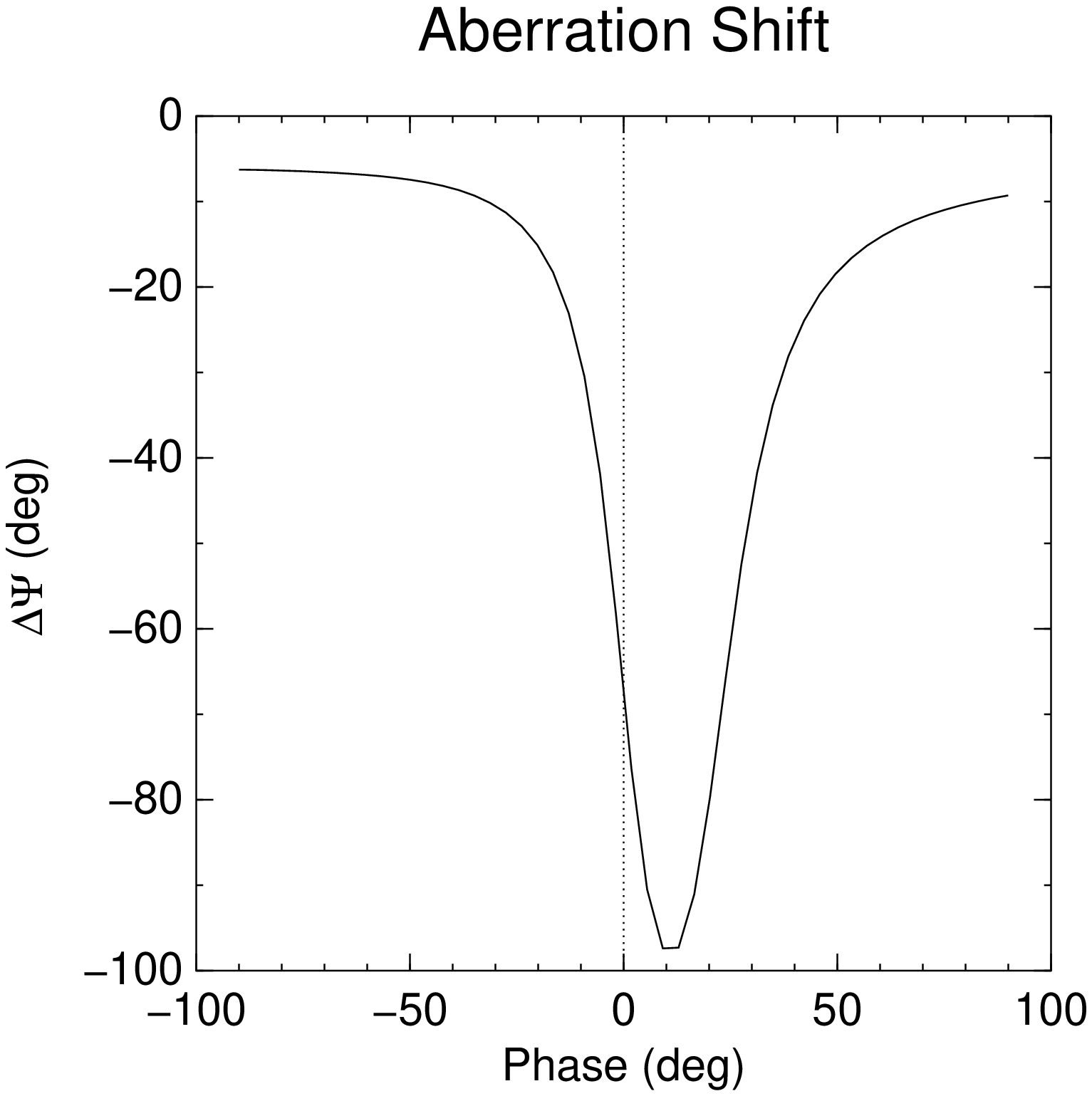}
  \caption{Perturbation to polarization angle, $\Delta\psi$, due to
  co-rotation effects as a function of longitude.  All angles are in
  degrees.  The dotted line is the unperturbed sweep, while the solid
  line is the perturbed sweep. $r/R_L = 0.1$, $\alpha = 30^\circ$, and
  $i=35^\circ$.
% The solid line is the numerical result, the dotted the theoretical.
\label{fig:aber}}
\end{figure}

\begin{figure}
  \plottwo{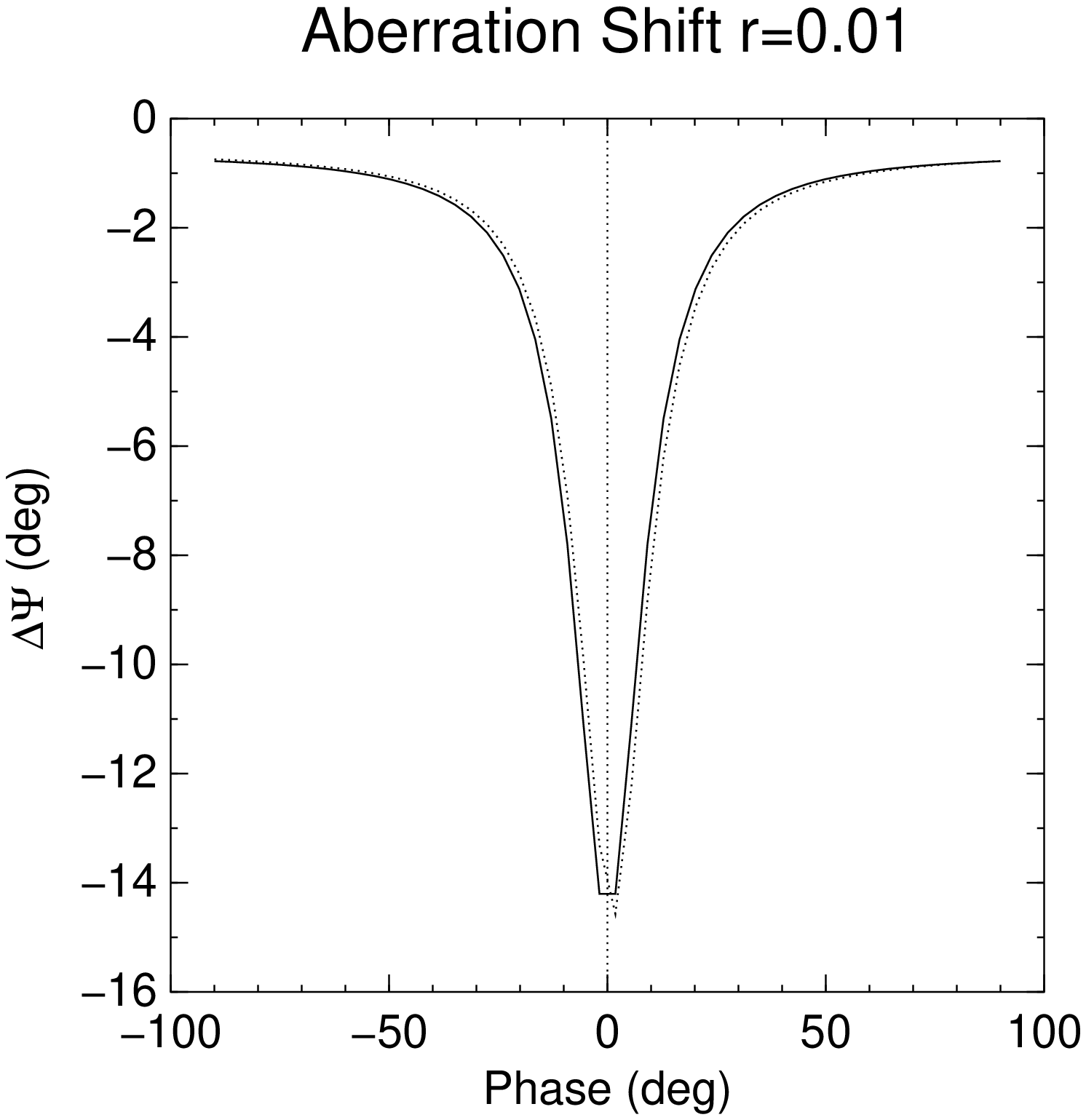}{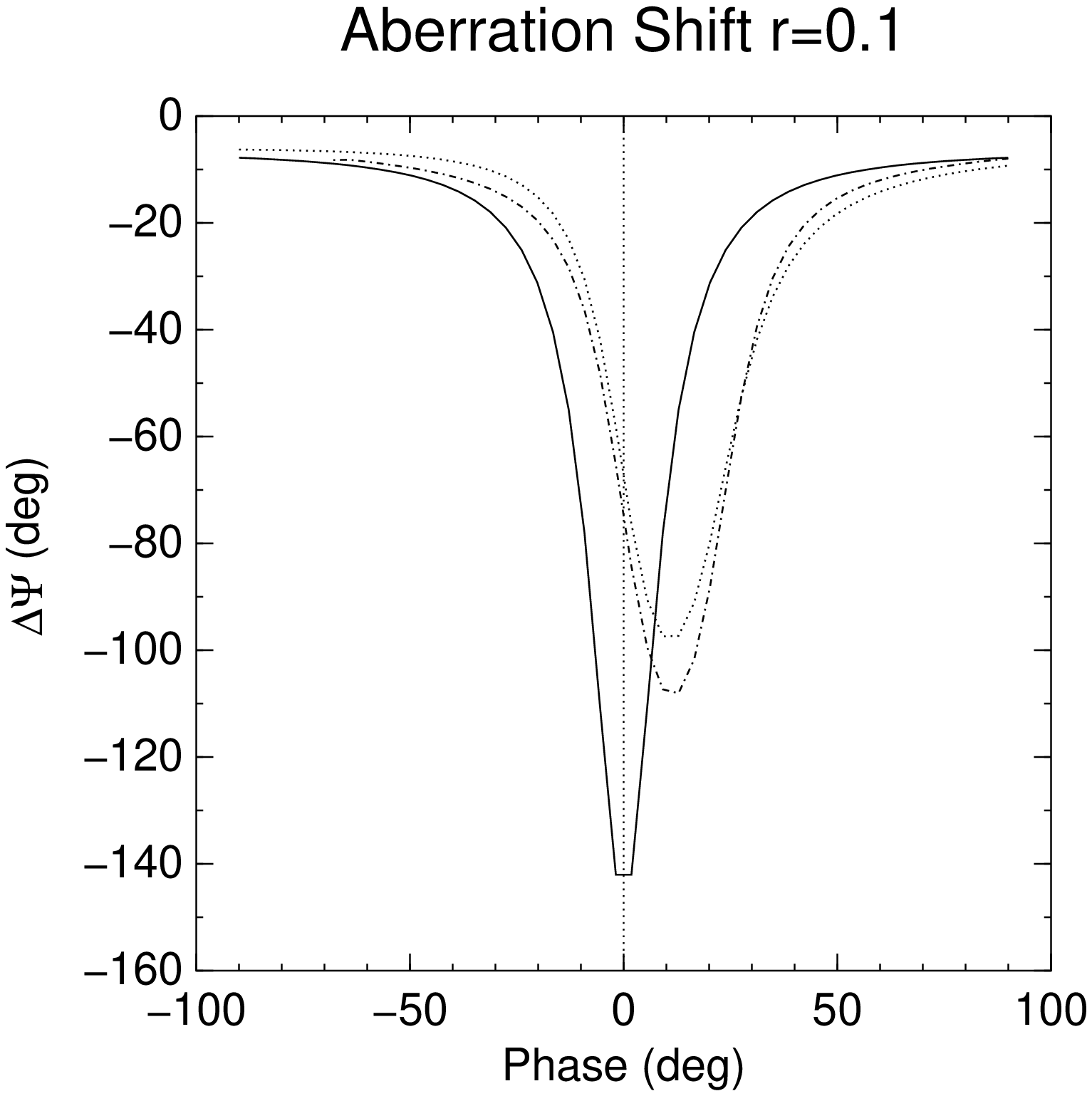}
  \caption{Analytic polarization angle shift due to aberration compared
to the numerical for $r/R_L = 0.01$ and $r/R_L = 0.1$.  $\alpha =
30^\circ$, and $i=35^\circ$.  The solid line is the analytic
perturbation theory, the dot-dash line (for $r/R_L = 0.1$) is the
phase shift derived from the perturbation theory, and the dotted line
is the numerical result.  Casting the perturbation as a phase shift
better matches the numerical result for large $r/R_L$.
% The solid line is the numerical result, the dotted the theoretical.
\label{fig:aber_diff}}
\end{figure}

If Figure \ref{fig:aber}, we show the numerically-calculated
polarization sweep for $r/R_L = 0.1$; this gives a good idea of the
character of the perturbation.  As shown in Figure
\ref{fig:aber_diff}, when $r/R_L = 0.01$, the analytically predicted
first-order shift is nearly identical to the numerical, while at
$r/R_L = 0.1$ there is a noticeable difference between the two, which
is mostly corrected by casting the perturbation as a phase shift.

Once this phase shift is removed, we are left with a remaining
perturbation of
\begin{equation}
  \Delta \Psi = - \frac{10}{3} \frac{r}{R_L} \cos \alpha \left(1 -
    \frac{7}{10} \sin^2 \theta_0 \right) +
    \frac{47}{12} \frac{r}{R_L} \sin \alpha \sin \theta_0 \cos \phi_0 +
    O(\sin^3 \theta_0).
\end{equation}
See Appendix \ref{ap:phase} for details.  In terms of the pulsar
phase, this is approximately
\begin{equation}
  \Delta \Psi \approx -\frac{10}{3} \frac{r}{R_L} \cos \alpha
    \left(\frac{3}{8} + \frac{5}{8} \cos \beta \right) +
    \frac{47}{18} \frac{r}{R_L} \sin \alpha \sin \beta -
    \frac{23}{144} \frac{r}{R_L} \sin 2 \alpha \sin i
      \sin^2 \Omega t
  \label{eq:aber_approx}
\end{equation}
using the same approximations as for Eq. (\ref{eq:current-approx}).

The constant portion of this polarization angle shift cancels the
constant portion of the perturbation due to polar current flow when
$J/J_{GJ} = 1$.  However, if $J/J_{GJ} \neq 1$, there is a nonzero
shift, providing a diagnostic of the magnitude of current flow.  The
remaining linear and quadratic terms produce slight changes to the
shape of the polarization angle sweep.

\subsection{Relativistic Sweep-back}

The actual magnetic field of a pulsar will differ from the ideal
dipole due to the relativistic sweep-back of field lines, caused by
the induced displacement current.  The full vacuum fields are given in
\citet{deutsch} and most recently restated by \citet{melatos}.
Admittedly, the region surrounding a pulsar is almost certainly not a
vacuum, but the morphology of the vacuum field should roughly reflect
the structure of the magnetic field with conduction currents.

The difference between the Deutsch fields and the vacuum dipole is
{\em second} order in $r/R_L$, however, so the first order effects
considered here dominate at low altitudes ($r/R_L < 0.2$).

In the non-vacuum case, the corresponding perturbation is the magnetic
field due to the rotation of the Goldreich-Julian density.  This field
component satisfies $\nabla \times \vc{B}_{rot} = 4 \pi (\vc{\Omega}
\times \vc{r}/c) \eta_{GJ}$.  Since both $\vc{\Omega} \times \vc{r}/c$
and $r \eta_{GJ}/B_0$ are manifestly first-order in $r/R_L$,
$B_{rot}/B_0$ is second order and therefore has been neglected.

\section{Return Currents}

While the perturbations described in the previous section are general
effects, this section applies those effects to a specific model of the
polar cap.  To maintain charge balance, the current flowing out from
the polar cap must be balanced by a return current elsewhere in the
system.  Here, we assume this current flows along the boundary layer
of last-closed field lines, surrounding the polar cone (Goldreich and
Julian 1969).

Outside this auroral sheath, the net current flow through a loop
surrounding the pole is zero, so there is no magnetic field
perturbation due to the current.  Only the aberrational shift remains.
The field on the open field lines stays the same, containing
perturbations from both current and aberration, while within the layer
itself the current perturbation is quickly eliminated.  For more
detailed analytic models, see Appendix \ref{ap:gen-current}.

\begin{figure}
  \plotone{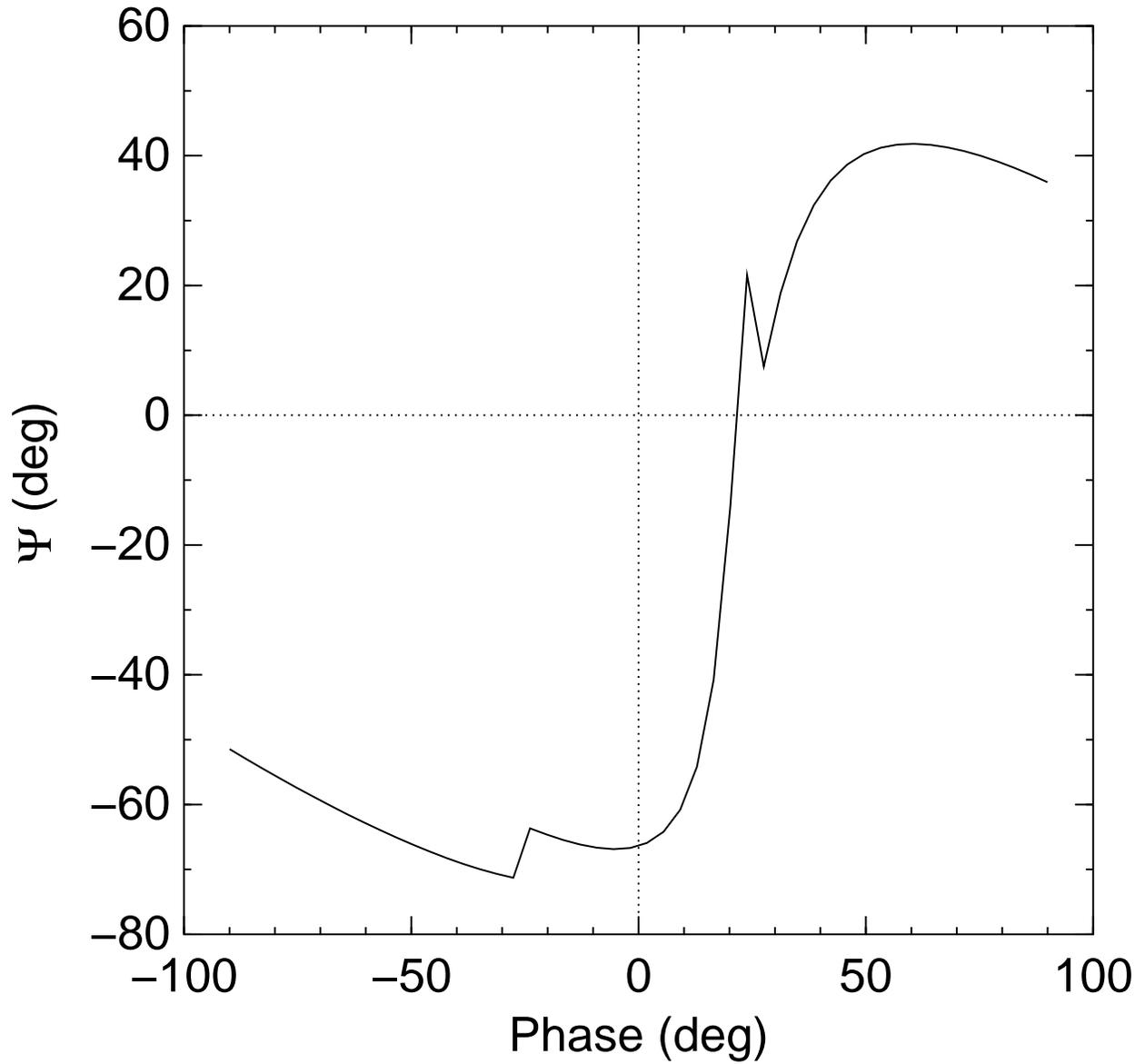}
  \caption{Polarization sweep for $\alpha=30^\circ$, $i=35^\circ$, and
    $r/R_L = 0.1$, with a return current layer at $\Phi = \pm
    25^\circ$ with a thickness set by $\lambda = 50$, showing the
    resultant jumps in the polarization angle sweep.
    \label{fig:combo}}
\end{figure}

If this return current layer is illuminated by emission, it will
produce a characteristic signature in the polarization sweep, Figure
\ref{fig:combo}.  The sharp transitions are caused by the current
perturbation turning on and off as the line of sight passes through
the return current layer.

Although the return current layer is not normally thought of as a site
of emission, it may be visible either through direct emission, by some
form of two-stream instability, or through refraction or scattering of
radiation emitted within the cone.  Since we are more interested in
the consequences of current flow above polar caps, we leave the
precise mechanism of the emission open.  If the return current is
composed of outflowing positrons or ions, the beaming properties would
be much the same as those of the primary beam, while if the return
current interpenetrates the outflow, even the identical mechansism
would reflect the presence of the return current.  Transitions like
the one in Figure \ref{fig:combo} would be strong evidence for
emission in this region or radiation transfer through it.

In addition to these specific effects, a more general conclusion is
that inhomogeneities on the order of $J_{GJ}$ in the current flow
above pulsar polar caps should produce perturbations in the
polarization angle sweep on the order of $r/R_L$.

\section{Discussion}

Given a single polarization angle sweep, at a single frequency,
observers can hope to see the effects of any current inhomogeneities,
small deviations from the standard shape caused by higher-order terms,
and the phase shift between the center of the polarization angle sweep
and the pulse itself.  The constant offsets caused by current flow
would be detectable only through comparison of multiple frequencies,
which would also help confirm any of the other effects.

From a more qualitative point of view, these concepts may be used to
help understand certain difficult objects.  For example, the
potentially large jumps in the polarization angle when the line of
sight crosses a return current layer could explain some of the shifts
which are normally ascribed to orthogonal modes.  Clearly, this result
does not explain orthogonal modes in general, as can be seen from the
single-pulse studies of XXX, but these shifts could masquerade as
orthogonal mode transitions in individual pulsars.

For example, in J0437-0415, the magnitude of the ``orthogonal''
transitions changes with frequency, ranging from $90^\circ$ at high
frequency to $70^\circ$ and $40^\circ$ at lower frequencies,
suggesting that either these are orthogonal modes combined with
propagation effects or, perhaps, that the $90^\circ$ shift is a
coincidence and these shifts are instead due to current features in
the magnetosphere.

In addition to polarization jumps, J0437-0415 displays a polarization
phase lag in which higher frequencies lag lower \citep{navarro}.
Combined with the increasing transition amplitude with frequency, this
suggests that the higher frequencies arise from higher in the
magnetosphere than the lower frequencies, contrary to the normal
radius-to-frequency map.  Since most millisecond pulsars do not
display the pulse widening with decreased frequency that is one of the
best arguments for radius-to-frequency mapping in normal pulsars, it
is certainly conceivable that the structure of the emission region is
different.  In any case, the small range of phase shifts seen argues
for a small emission region, on the order of a few kilometers.

\section{Summary}

In this paper, we have shown how two physical effects, rotational
aberration and current flow, affect the polarization angle of observed
radiation.  The strength of these effects is proportional to the
height of the emission region.

In brief, current flow tends to shift the entire polarization angle
sweep upwards in a plot of polarization angle vs. phase, while
aberration tends both to shift the sweep downwards and to phase shift
it.  The largest of these is the aberrational phase shift between the
center of the emission profile and the center of the polarization
sweep.

The Goldreich-Julian current and aberration produce equal and opposite
constant offsets to the polarization angle sweep, effectively
cancelling each other.  Any deviations in the current from the
fiducial Goldreich-Julian value will break this cancellation, giving a
potential indicator of the magnitude of current flow above the polar
cap.

Inhomogeneities in the current flow structure above the polar cap
should be revealed as perturbations, of order $r/R_L$, in the
polarization sweep.  The specific case of an auroral return current
layer would appear as sharp transitions along the edge of the observed
pulse, potentially correlated with conal emission.  These jumps may
appear similar to orthogonal modes, but should have an amplitude which
changes with frequency, if frequency is related to emission radius.
In the presence of current sheets, single-pulse polarimetry should
show a shift in both orthogonal modes, provided that multiple
altitudes do not contribute to the observed radiation at one
frequency.  This is quite different from the existing single-pulse
studies \citep{gil95, ganga97}, which show a simple shift in the relative
amplitude of two existing modes.

\acknowledgements

We would like to thank Don Backer and Andrea Somer for illuminating
discussions about the observability of these effects and Anatoly
Spitkovsky for helpful animations of swept-back field lines.

\appendix

\section{Geometric Method}
\label{ap:method}

From classical differential geometry, the tangent vector is the unit
vector along $\vc{B}$, $\uvc{t} = \vc{B}/|B|$, while the normal vector
$\uvc{n}$ is the unit vector along the radius of curvature.  For
highly relativistic flow, radiation is beamed along the direction of
the particle velocity, so at any instant the observed radiation comes
from the spot in the magnetosphere where the tangent vector $\uvc{t}$
points in the direction of the observer.

For normal vacuum radiation, the polarization vector must lie in the
plane perpendicular to $\uvc{t}$; for the particular case of curvature
radiation, the polarization is almost entirely along $\uvc{n}$.  For
plasma processes, this restriction is relaxed, but in the absence of
other directional perturbations, such as density gradients not along
$\uvc{n}$, we expect that the polarization angle would be related to
the direction of curvature of the field lines.
Density gradients along $\uvc{n}$, which we expect if polar cap pair
creation is the main source of the plasma, would likewise link the
polarization to the magnetic geometry.

To trace the orientation of the field, we use the binormal vector,
defined by $\vc{b} \equiv \uvc{t} \times \uvc{n}$.  This is equivalent
to tracing the normal, but is computationally more convenient,
since the binormal of a dipole field is simply a constant vector along
$\bvc{\phi}$.

The observed polarization angle is then set by finding the angle
between the normal vector and the projected rotation axis of the
pulsar.  Since the tangent points towards the observer, both the
normal and the binormal are automatically in the plane of the sky, but
the rotation axis must be projected, giving $\vc{\Omega}_p \equiv
\uvc{\Omega} - (\uvc{\Omega} \cdot \uvc{t}) \uvc{t}$.

If the angle between the rotation axis and the binormal is $\psi'$,
then $\cos \psi' = \uvc{b} \cdot \uvc{\Omega}_p$ and $\sin \psi' =
\uvc{t} \cdot (\uvc{b} \times \uvc{\Omega}_p)$.  Since the angle made
by the normal vector is $90^\circ$ different from that made by the
binormal, the true polarization angle is given by
\begin{equation}
  \tan \psi = -\cot \psi' = \frac{\uvc{b} \cdot \uvc{\Omega}}{\uvc{b}
    \cdot (\uvc{t} \times \uvc{\Omega})}
  \label{eq:tanpsi-basic}
\end{equation}
where we have expressed the result in terms of $\uvc{\Omega}$, rather
than $\vc{\Omega}_p$.

For a standard dipole field, this gives the normal polarization curve,
\begin{equation}
  \tan \psi = \frac{\sin \alpha \sin \Omega t}{\cos \alpha \sin i -
    \sin \alpha \cos i \cos \Omega t},
  \label{eq:RC}
\end{equation}
c.f. \citet{rc}.

In general, we want to know the change in the polarization angle due
to a small perturbation to the background magnetic field, $\vc{B} =
\vc{B}_0 + \vc{B}_1$.  This change is due to shifts in both the
emission point and the binormal itself.

To first order, the new tangent vector $\uvc{t}$ will be the sum of
the initial tangent vector, $\uvc{t}_0$, and a perturbation $\vc{t}_1$
perpendicular to $\uvc{t}_0$.  (Any component along $\uvc{t}_0$ is
simply absorbed into the normalization.)

The new curvature vector is
\begin{equation}
  \vc{\kappa} = (\uvc{t} \cdot \nabla) \uvc{t} =
    (\uvc{t}_0 \cdot \nabla) \uvc{t}_0 + (\uvc{t}_0 \cdot \nabla) \vc{t}_1
    + (\vc{t}_1 \cdot \nabla) \uvc{t}_0 = \vc{\kappa}_0 + \vc{\kappa}_1
\end{equation}
where $\vc{\kappa} \equiv \uvc{n}/\rho$, $\vc{\kappa}_0 =
\uvc{n}_0/\rho_0 = (\uvc{t}_0 \cdot \nabla) \uvc{t}_0$, and
\begin{equation}
  \vc{\kappa}_1 = (\uvc{t}_0 \cdot \nabla) \vc{t}_1
    + (\vc{t}_1 \cdot \nabla) \uvc{t}_0
  \label{eq:kappa_1}
\end{equation}
When normalized, the component of $\vc{\kappa}_1$ along $\uvc{n}_0$ is
absorbed into the normalization, giving a perturbed normal vector of
$\uvc{n} = \uvc{n}_0 + \vc{n}_1$, where
\begin{equation}
  \vc{n}_1 = \rho_0 (\vc{\kappa}_1 - (\uvc{n}_0 \cdot \vc{\kappa}_1)\uvc{n}_0).
  \label{eq:perturbed-n}
\end{equation}

The perturbed binormal is then $\uvc{b} = \uvc{b}_0 + \vc{b}_1$, where
\begin{equation}
  \vc{b}_1 = \vc{t}_1 \times \uvc{n}_0 + \uvc{t}_0 \times \vc{n}_1.
\end{equation}

The second contribution to the polarization angle change comes from
the change in the emission point.  If the unperturbed field points
towards the observer at $\vc{r}_0$, the perturbed does so at $\vc{r}_0
+ \vc{r}_1$, where $\vc{r}_1$ can be found by solving
\begin{equation}
  \uvc{t}_0(\vc{r}_0) + \nabla \uvc{t}_0 \cdot \vc{r}_1 +
    \vc{t}_1(\vc{r}_0) = \uvc{\ell}
\end{equation}
or
\begin{equation}
   \vc{r}_1 = -[\nabla \vc{t}_0]^{-1} \cdot \vc{t}_1(\vc{r}_0)
\end{equation}
This expression is underdetermined, as there is typically an entire
line of points where the field points towards the observer.

In this paper, we assume that emission at a given frequency arises at
(or decouples from the magnetic field at) a fixed radius, so we
constrain the radius to remain constant.  This ``lateral'' motion
shifts the emission site to a different field line from the
unperturbed one, and, if transverse gradients are steep, with a size
scale shorter than roughly $r^2 \sin \theta/R_L$, this will likely
affect the emission properties.  Near the boundaries of the polar cap,
such transverse gradients may become important, requiring a more
careful treatment.

This changes the observed binormal by an additional $\vc{\delta b} =
\nabla \uvc{b}_0 \cdot \vc{r}_1$, giving a total change of $\Delta
\vc{b} = \vc{b}_1 + \vc{\delta b}$.  Since this change in the binormal
is perpendicular to the unperturbed binormal, the magnitude of the
change in polarization angle is simply $|\Delta \psi| = |\Delta
\vc{b}|$.

Getting the sign correct requires comparing the change in the binormal
to the direction of the projected rotation axis, or
\begin{equation}
  \Delta \psi = \frac{\uvc{t}_0 \cdot (\Delta \vc{b} \times
    \Omega_p)}{\uvc{b}_0 \cdot \Omega_p}.
\end{equation}

Since all of these steps are linear in the perturbation, multiple
perturbations may be considered separately, then simply added.  In
later sections, we consider in turn two different possible
perturbations to the basic dipole field.

\section{Unperturbed Field}
\label{ap:unperturbed}

In $(r, \theta, \phi)$ polar coordinates centered on the magnetic
axis, the unperturbed dipole field is
\begin{equation}
  \vc{B}_0 = \left[\frac{2 \mu \cos \theta}{r^3},\,
                \frac{\mu \sin \theta}{r^3},\, 0 \right]
\end{equation}
which gives a tangent vector field of
\begin{equation}
  \uvc{t}_0 \equiv \frac{\vc{B}_0}{|B_0|} = \frac{1}{N_1}
  \left[\cos \theta,\, \frac{1}{2} \sin \theta, \, 0\right]
\end{equation}
where
\begin{equation}
   N_1 \equiv \left( 1 - \frac{3}{4} \sin^2 \theta \right)^{1/2}
   \label{eq:n1}
\end{equation}

The curvature vector is then
\begin{equation}
  \vc{\kappa}_0 \equiv % \nabla_{\uvc{t}_0} \uvc{t}_0 = 
    (\uvc{t}_0 \cdot \nabla) \uvc{t}_0 =
    \frac{3}{4} \frac{\sin \theta}{r N_1^4} (1 - \frac{1}{2} \sin^2
    \theta) \left[-\frac{1}{2} \sin \theta, \, \cos \theta, \,
    0\right],
\end{equation}
giving a normal vector of
\begin{equation}
  \uvc{n}_0 \equiv \frac{\vc{\kappa}_0}{|\kappa_0|} =  \frac{1}{N_1}
  \left[ -\frac{1}{2} \sin \theta, \, \cos \theta, \, 0\right].
\end{equation}

The binormal is purely in the $\phi$-direction,
\begin{equation}
  \uvc{b}_0 = \uvc{t}_0 \times \uvc{n}_0 = [0, \, 0, \, 1].
\end{equation}

In order to relate this binormal field to the actual polarization
angle, we need to know the emission point, which we denote $\vc{r}_0 =
(r_0,\, \theta_0,\, \phi_0)$, where the magnetic field points towards
the observer at $(\theta_{obs},\, \phi_{obs})$.

Since the dipole field contains no $\phi$-component, we must have
$\phi_0 = \phi_{obs}$.  Solving for $\theta_0$ then gives
\begin{equation}
  \tan \theta_0 = \frac{3}{2} \frac{1}{\tan \theta_{obs}}
    \left\{ \pm \left(1 +
    \frac{8}{9} \tan^2 \theta_{obs} \right)^{1/2} - 1 \right\}
  \label{eq:tantheta0}
\end{equation}
If $\theta_{obs}$ is small, as is usually the case for an observer
looking down the cone of open field lines, this simplifies to
$\theta_0 \approx (2/3) \theta_{obs}$.

If we take these results and evaluate expression
(\ref{eq:tanpsi-basic}), we obtain the standard polarization angle
sweep, (\ref{eq:RC}).

\section{Approximations}
\label{ap:approximations}

Evaluating the polarization effects of various perturbations naturally
leads to expressions which depend on the coordinates of the emission
point, usually in the frame co-rotating with the star.  Since no
actual observer is sitting in that co-rotating frame, we have to
rewrite the results in terms of observables, such as the rotational
phase ($\Omega t$) and the physical inclination angles ($\alpha,\,
i$).  Admittedly, the inclination angles are not directly observable,
but they set the geometry of the pulsar.

Unfortunately, this conversion is a significant source of error
in the analytic perturbation theory.  Straightforward power-series
expansions may be used to cast the raw analytical expressions into
more useful forms, but doing so limits the range of validity to a
small portion of the pulse phase near the pulse center.

Here, we first give the exact formulae relating the pulse phase to the
position of the observer in these coordinates.  These, coupled with
the solution for the 0-order emission colatitude (\ref{eq:tantheta0}),
give a precise definition for $\theta_0$ and $\phi_0$ at any pulse
phase.  We then give several more tractable, but more limited,
approximations.

The vector to the observer is
\begin{equation}
  \uvc{\ell} = \left[
    \begin{array}{c}
      \cos \alpha \sin i \cos \Omega t - \sin \alpha \cos i \\
      - \sin i \sin \Omega t \\
      \sin \alpha \sin i \cos \Omega t + \cos \alpha \cos i
    \end{array}
    \right]
\end{equation}
in Cartesian magnetic coordinates.  In polar, the observer is at
$(\theta_{obs}, \phi_{obs})$ given by
\begin{eqnarray}
  \cos \theta_{obs} & = & \cos \alpha \cos i + \sin \alpha \sin i \cos
    \Omega t \label{eq:cos_th_obs} \\
  \tan \phi_{obs} & = & \frac{\sin i \sin \Omega t}{\sin \alpha \cos i -
    \cos \alpha \sin i \cos \Omega t}.
    \label{eq:tan_ph_obs}
\end{eqnarray}
These expressions, coupled with (\ref{eq:tantheta0}) and the fact that
$\phi_0 = \phi_{obs}$, fully define the emission coordinates.

For emission from within the cone of open field lines, several
of these angles are small.  If the observer is to see this cone, the
difference between $\alpha$ and $i$ must be less than (or on the order
of) the colatitude of the last closed field line, $\theta_c(r) =
\sqrt{r / R_L}$.  Similarly, the magnetic colatitude of the emission,
$\theta_0$, and the rotational phase $\Phi = \Omega t$ must be of this
same order.

The expression for $\tan \theta_0$ (\ref{eq:tantheta0}) is perhaps the
most difficult formula.  It has two roots, and the continuous physical
solution switches between them when $\theta_{obs} = \pi/2$.  Since the
maximum of $\theta_{obs}$ is $\alpha + i$, this switch will occur
somewhere in the pulse phase for all pulsars with $\alpha + i >
\pi/2$.  If we are only interested in the region close to the pulse
center, then this is not a problem, but for millisecond pulsars and
other pulsars whose emission covers a substantial fraction of the
period, this switch-over has to be kept in mind.

Here, however, for simplicity we assume that $\theta_{obs}$ is always
less than $\pi/2$.  In that case, an excellent approximation to
(\ref{eq:tantheta0}) is
\begin{equation}
  \tan \theta_0 \approx \frac{3}{2} \frac{1}{\sin \theta_{obs}}
    \left(1 - \cos \theta_{obs} - \frac{1}{18} \sin^2 \theta_{obs}
    \right)
\end{equation}
while for small $\theta_{obs}$,
\begin{equation}
  \tan \theta_0 \approx \sin \theta_0 \approx \frac{2}{3} \sin
    \theta_{obs}
\end{equation}

Expressing $\sin \theta_{obs}$ in terms of $\cos \theta_{obs}$ gives
us, for small $\theta_{obs}$,
\begin{equation}
  \sin \theta_{obs} = \left[ 2 (1 - \cos \theta_{obs}) \right]^{1/2}
\end{equation}
which is quite useful, since we frequently have terms in even powers
of $\sin \theta_0$, and we know $\cos \theta_{obs}$ from
Eq. (\ref{eq:cos_th_obs}).

The observer and emission azimuth are the same, $\phi_{0} =
\phi_{obs}$.  Using the transformed y-coordinate, $y_B = r \sin \theta
\cos \phi$, we find
\begin{eqnarray}
  \cos \phi_0 & = & \frac{\cos \alpha \sin i \cos \Omega t - \sin \alpha
    \cos i}{\sin \theta_{obs}} \\
  \sin \phi_0 & = & \frac{-\sin i \sin \Omega t}{\sin \theta_{obs}}
\end{eqnarray}
By using the relations between $\sin \theta_0$ and $\sin
\theta_{obs}$, these can be cast into various useful forms.

For example, in order to simplify expressions
(\ref{eq:current-approx}) and (\ref{eq:aber_approx}), we used these
relations to write
\begin{equation}
  \sin^2 \theta_0 \approx \frac{4}{9} \sin^2 \theta_{obs} \approx
    \frac{4}{9} \left(1 - \cos \beta +
      \frac{1}{2} \sin \alpha \sin i \sin^2 \Omega t \right)
\end{equation}
and
\begin{equation}
  \sin \theta_0 \cos \phi_0 \approx \frac{2}{3} \sin \theta_{obs}
    \cos \phi_{obs}
    \approx \frac{2}{3} \left(\sin \beta -
      \frac{1}{2} \cos \alpha \sin i \sin^2 \Omega t \right)
\end{equation}
where $\beta \equiv i - \alpha$ and we have taken $\cos \Omega t
\approx 1 - (1/2) \sin^2 \Omega t$.  As mentioned before, this limits
these expressions to apply only to the region close to the pole where
$\theta_{obs}$ is small.

\section{Field Aligned Current}
\label{ap:current}

We expect a current to flow along the open field lines with intensity
approximately equal to the Goldreich-Julian charge density moving at
the speed of light,
\begin{equation}
  \vc{J} = -\frac{1}{2 \pi} \zeta (\vc{\Omega} \cdot \vc{B}) \uvc{B}
\end{equation}
where $\zeta = J/J_{GJ}$ is scale factor of order unity, reflecting
our ignorance of the actual current.  Since we will be examining only
first-order effects, the results for the perturbing fields and the
polarization shifts are simply proportional to $\zeta$.  In the
interest of conciseness, we set $\zeta=1$ for the rest of this
section and reinsert it into the final result.

In the case of a small polar cap, the variation in $\vc{\Omega} \cdot
\vc{B}$ across the cap is also small, so we take $\vc{\Omega} \cdot
\vc{B} \approx \Omega B \cos \alpha$.  In addition, if $\vc{B} = \vc{B}_0 +
\vc{B}_1$, with $B_1 \ll B_0$ and $\nabla \times \vc{B}_0 = 0$, as is
true for any superposition of vacuum multipoles, then $\nabla \times
\vc{B}_1 = 2 \Omega_* {\bf B}_0 /c$.

Now assume $\vc{B}_0$ is a point dipole and introduce spherical polar
coordinates with the $z$ axis along the dipole axis and the origin of
coordinates at the point dipole.  Assume axisymmetry with respect to
the magnetic axis.  Since the assumed current is axisymmetric, this
assumption is equivalent to assuming the polar flow tube has a
circular cross section.

The equations for the field components are
\begin{eqnarray}
   \frac{1}{r \sin \theta} \frac{\partial}{\partial \theta}
      (B_{1\phi}\sin \theta) & = & -\frac{4 \Omega_* \mu}{c r^3}
      \cos \alpha \cos \theta, \\
   - \frac{1}{r} \frac{\partial}{\partial r}(r B_{1\phi}) & = &
      -\frac{2 \Omega_* \mu}{c r^3} \cos \alpha \sin \theta, \\
   \frac{1}{r} \frac{\partial}{\partial r}(r B_{1\theta}) - 
      -\frac{1}{r}\frac{\partial B_{1r}}{\partial \theta} = 0.
\end{eqnarray}

Here $B_{0 r} = (2\mu /r^3) \cos \theta , \; B_{0\theta} = (\mu
/r^3) \sin \theta$.  The solution is
\begin{eqnarray}
   B_{1r} & = & B_{1\theta} =0, \\
   B_{1\phi} & = & -\frac{2\mu}{R_L^3} \cos \alpha
                   \left( \frac{r}{R_L} \right)^{-2} \sin \theta,
\end{eqnarray}
where $R_L \equiv c/\Omega_*$.  One readily finds this $B_{1\phi}$
satisfies both of the non-trivial equations.

This gives a perturbation to the tangent vector of
\begin{equation}
  \vc{t}_1 = \left[0,\,0,\, -\frac{r}{R_L} \cos \alpha
    \frac{\sin \theta}{N_1}\right]
\end{equation}
in $(r,\, \theta,\, \phi)$ polar components.

Following the above method, and expanding in powers of $\sin \theta$,
the updated normal is
\begin{equation}
  \vc{n}_1 = \left[0,\, 0,\, -4 \frac{r}{R_L} \cos \alpha (1 - \frac{1}{4} \sin^2
    \theta) \right]
\end{equation}
which results in a change in the binormal of
\begin{equation}
  \vc{b}_1 = \frac{r}{R_L} \cos \alpha \left[-\sin \theta,\,
        4 (1 - \frac{1}{4} \sin^2 \theta),\, 0 \right].
\end{equation}

The emission point changes by
\begin{equation}
  \vc{r}_1 = -[\nabla \uvc{t}_0]^{-1} \cdot \vc{t}_1 = \frac{r^2}{R_L}
    \cos \alpha 
    \left[0,\, 0,\, \frac{2}{3} N_1 \tan \theta \right],
\end{equation}
changing the observed binormal by
\begin{equation}
  \delta \vc{b} = \nabla \uvc{b}_0 \cdot \vc{r}_1 =
    -\frac{r}{R_L} \cos \alpha \left[
    \frac{2}{3} \sin \theta,\,
    \frac{2}{3},\, 0 \right]
\end{equation}
giving a total perturbation of
\begin{equation}
  \Delta \vc{b} = \vc{b}_1 + \delta \vc{b} = \frac{r}{R_L} \cos \alpha \left[
    -\frac{5}{3} \sin \theta,\, \frac{10}{3} - \sin^2 \theta \right]
\end{equation}
yielding a polarization angle shift of
\begin{equation}
  \Delta \psi = \frac{10}{3} \frac{r}{R_L} \frac{J}{J_{GJ}} \cos \alpha
      \left(1 - \frac{7}{40} \sin^2 \theta \right),
\end{equation}
where we have re-inserted the dependence on the current strength.

\section{Return Current}
\label{ap:gen-current}

To model a return current, we assume that each field line has a fixed
flux of current on it, proportional to the local GJ density.

If the perturbing current density has the form
\begin{equation}
  \vc{J} = -\frac{1}{2 \pi} \Omega \cos \alpha \vc{B}
    f(\frac{\sin^2 \theta}{r})
\end{equation}
for $f$ an arbitrary function, then the perturbing field is
\begin{equation}
  B_{1\phi} = -\frac{2 \Omega \mu}{c} \cos \alpha \frac{1}{r \sin
  \theta} \int_0^{u} du'\, f(u')
\end{equation}
where $u \equiv \sin^2 \theta/r$ labels an individual field line.
This is equivalent to using the magnetic flux as the integration
variable.

Dividing by the magnitude of the field gives
\begin{equation}
  \vc{t}_1 = \frac{\vc{B}_1}{|B_0|} = - \frac{\cos \alpha}{N_1} \frac{r}{R_L}
        \frac{r}{\sin \theta} \int_0^{u} du'\, f(u') \: \bvc{\phi}
\end{equation}
with $N_1$ as defined in (\ref{eq:n1}).

If $f(u) = 1$, we get the normal perturbing field.  If we include a
return current which is a multiple $\lambda$ of the background GJ
current in a small ring past $\sin^2 \theta/r = u_1$, namely
% \begin{equation}
%   f(u) = \left\{
%     \begin{array}{cr}
%       1, & u_1 > u > 0 \\
%       -\lambda, & (1+1/\lambda) u_1 > u > u_1 \\
%       0, & u > (1+1/\lambda) u_1
%     \end{array}
%   \right.
% \end{equation}
\begin{equation}
  f(u) = \cases{1,&if $u_1 > u > 0$;\cr
                -\lambda,&if $(1+1/\lambda) u_1 > u > u_1$;\cr
                0,&if $u > (1+1/\lambda) u_1$;\cr}
\end{equation}
we get a perturbation of
% \begin{equation}
%   \vc{t}_1 = \frac{r}{R_L} \cos \alpha \; \bvc{\phi} \left\{
%     \begin{array}{cr}
%       - \sin \theta, & u_1 > u > 0 \\
%       - \frac{(1+\lambda) r u_1}{\sin
%           \theta} + \lambda \sin \theta, & 
%           (1+1/\lambda) u_1 > u > u_1 \\
%       0, & u > (1+1/\lambda) u_1
%     \end{array}
%   \right.
% \end{equation}
\begin{equation}
  \vc{t}_1 = \frac{r}{R_L} \cos \alpha \; \bvc{\phi} \cases{
        - \sin \theta,&if $u_1 > u > 0$ \cr
        - \frac{(1+\lambda) r u_1}{\sin
            \theta} + \lambda \sin \theta, &if
            $(1+1/\lambda) u_1 > u > u_1$ \cr
        0, &if $u > (1+1/\lambda) u_1$\cr}
\end{equation}

This generates the normal polarization angle shift for $u < u_1$, no
shift for $u > (1+1/\lambda) u_1$, and a shift of
\begin{equation}
  \Delta \psi = \frac{10}{3} \frac{r}{R_L} \lambda \cos \alpha \left(
    \frac{1+\lambda}{\lambda} \frac{u_1 r}{\sin^2 \theta} - 1 \right)
    \left(1 - \frac{7}{40} \sin^2 \theta \right)
%  \Delta \psi = r \cos \alpha \left\{ 2 (1+\lambda) \frac{r
%  u_1}{\sin^2 \theta} \left(1 - \frac{1}{24} \sin^2 \theta \right) -
%  \frac{10}{3} \lambda \left(1 - \frac{7}{40} \sin^2 \theta \right).
\end{equation}
for $(1+1/\lambda)u_1 > u > u_1$.  If $\lambda \gg 1$, then this
reduces to
\begin{equation}
  \Delta \psi = -\frac{10}{3} \frac{r}{R_L} \cos \alpha (1 - \xi)
    \left(1 - \frac{7}{40} \sin^2 \theta \right)
\end{equation}
where $\xi \equiv \lambda (u-u_1)/u_1$ ranges from 0 to 1 through the
return current layer.

This simply eliminates the perturbation over the range from
$(1+1/\lambda) u_1 > u > u_1$, which is what we expected.  If the
return current layer is thin, $\lambda \gg 1$, this transition is
abrupt.  This would appear in the data as a constant negative shift
outside of the region of current flow, as there would no longer be any
current to offset the aberration, as shown in Figure \ref{fig:combo}.

\section{Aberration}
\label{ap:aberration}

The aberrational perturbation arises because the velocity of particles
streaming along field lines has components both along the rotation
direction and along the field line.  Following \citet{blas}, the
velocity of the particles becomes
\begin{equation}
  \vc{v} = c \beta_\parallel \uvc{B} + \vc{\Omega} \times \vc{r},
\end{equation}
where $\beta_\parallel$ is a free parameter, reflecting the fraction
of particle velocity along the field.  Assuming the particles are
moving at $c$ and normalizing gives a change in the tangent vector of
\begin{equation}
  \vc{t}_1 = \frac{\vc{\Omega} \times \vc{r}}{c} -
    \left(\frac{\vc{\Omega} \times \vc{r}}{c} \cdot \uvc{t}_0\right)
    \uvc{t}_0
\end{equation}

This change in the tangent vector is of the same form as the current
aberration treated previously, i.e. first-order in $r/R_L$ and
perpendicular to $\uvc{t}_0$.  As such we can apply the same general
formalism to calculate the expected polarization angle shift.
Additional first-order changes to the magnetic field generate
second-order changes in $\vc{t}_1$, so we may safely treat this
perturbation independently.

In polar coordinates,
\begin{equation}
  \uvc{\Omega} \times \uvc{r} = \left[0,\, \sin \alpha \sin \phi,\,
    \sin \alpha \cos \theta \cos \phi + \cos \alpha \sin \theta \right]
\end{equation}

Subtracting out the component along $t_0$ gives
\begin{equation}
  \vc{t}_1 = \frac{r}{R_L} \left[
    \begin{array}{c}
      -\frac{1}{2} N_1^{-2} \sin \alpha \sin \theta \cos \theta 
        \sin \phi \\
      N_1^{-2} \sin \alpha \cos^2 \theta \sin \phi \\
      \sin \alpha \cos \theta \cos \phi + \cos \alpha \sin \theta
    \end{array}
  \right].
\end{equation}

Aberrational effects change the curvature of particle orbits in two
ways.  Not only does the instantaneous velocity of the particles
change by $\vc{\Omega} \times \vc{r}$, but the underlying magnetic
dipole is rotating as well.  The adds a $\partial \vc{t}_0/\partial t$
term to the curvature vector perturbation $\vc{\kappa}_1$ from
(\ref{eq:kappa_1}), giving
\begin{equation}
  \vc{\kappa}_1 = \frac{\partial \vc{t}_0}{\partial t} +
    (\vc{t}_0 \cdot \nabla) \vc{t}_1 + (\vc{t}_1 \cdot \nabla) \vc{t}_0
\end{equation}

The normal vector is then found from (\ref{eq:perturbed-n}) to be
\begin{equation}
  \vc{n}_1 = \frac{r}{R_L} \left[
    \begin{array}{c}
      -\sin \alpha \sin \phi \left(1 - \frac{1}{4} \sin^2 \theta \right) \\
      -\frac{1}{2} \sin \alpha \sin \theta \sin \phi \\
      \frac{8}{3} \frac{\sin \alpha \cos \phi}{\sin \theta}
        \left(1 - \frac{7}{4} \sin^2 \theta \right) +
        4 \cos \alpha \left(1 - \frac{3}{4} \sin^2 \theta \right)
     \end{array}
    \right] + O(\sin^3 \theta)
\end{equation}
producing a binormal change of
\begin{equation}
  \vc{b}_1 = \frac{r}{R_L} \left[
    \begin{array}{c}
      \frac{1}{3} \sin \alpha \cos \phi \left(1 - \frac{29}{8} \sin^2
        \theta \right) + \cos \alpha \sin \theta \\
      -\frac{8}{3} \frac{\sin \alpha \cos \phi}{\sin \theta}
        \left(1 - \frac{27}{16} \sin^2 \theta\right)
        -4 \cos \alpha \left(1 - \frac{3}{4} \sin^2 \theta \right) \\
      0
    \end{array} \right] + O(\sin^3 \theta)
\end{equation}

Solving for the emission point gives
\begin{equation}
  \vc{r}_1 = \frac{r^2}{R_L} \left[
    \begin{array}{c}
      0 \\
      -\frac{2}{3} \sin \alpha \sin \phi
        \left(1 - \frac{3}{8} \sin^2 \theta \right) \\
      -\frac{2}{3} \sin \alpha \cos \phi \left(1 -
        \frac{3}{8} \sin^2 \theta \right) -
        \frac{2}{3} \cos \alpha \sin \theta
    \end{array} \right] + O(\sin^3 \theta)
\end{equation}
and a change in binormal of
\begin{equation}
  \delta \vc{b} = \frac{r}{R_L} \left[
    \begin{array}{c}
      \frac{2}{3} \sin \alpha \cos \phi
        \left(1 - \frac{3}{8} \sin^2 \theta \right) +
        \frac{2}{3} \cos \alpha \sin \theta \\
      \frac{2}{3} \frac{\sin \alpha \cos \phi}{\sin \theta}
        \left(1 - \frac{7}{8} \sin^2 \theta \right) +
        \frac{2}{3} \cos \alpha \\
      0
    \end{array} \right] + O(\sin^3 \theta)
\end{equation}
giving a total change in the binormal of
\begin{equation}
  \Delta \vc{b} = \frac{r}{R_L} \left[
    \begin{array}{c}
      \sin \alpha \cos \phi \left(1 - \frac{35}{24}
        \sin^2 \theta \right) + \frac{5}{3} \cos \alpha \sin \theta \\
      -2 \frac{\sin \alpha \cos \phi}{\sin \theta}
        \left(1 - \frac{47}{24} \sin^2 \theta \right) -
        \frac{10}{3} \cos \alpha \left(1 - \frac{33}{40}
          \sin^2 \theta \right) \\
      0
    \end{array} \right] + O(\sin^3 \theta)
    \label{eq:b1aberr}
\end{equation}
and a shift in the polarization angle of
\begin{equation}
  \Delta \psi = -2 \frac{r}{R_L} \frac{\sin \alpha \cos \phi}{\sin
    \theta} \left(1 - \frac{11}{6} \sin^2 \theta \right) -
    \frac{10}{3} \frac{r}{R_L} \cos \alpha \left(1 -
    \frac{7}{10} \sin^2 \theta \right)
    + O(\sin^3 \theta).
\end{equation}

\section{Aberrational Phase Shift}
\label{ap:phase}

Qualitatively, the changes to the polarization angle caused by
aberration appear to simply shift the phase of the sweep.  BCW
directly converted their calculated perturbation into a phase shift,
but here we want to be slightly more precise and preserve the
higher-order terms (in $\sin \theta$).  A portion of the aberrational
change in the binormal $\Delta \vc{b}$ (\ref{eq:b1aberr}) corresponds
to a phase shift, but part of it does not.  In order to separate these
two parts, we need to calculate how the binormal at the emission point
changes with time.

This will let us decompose the perturbation into $\Delta \vc{b} =
\Delta \Phi (\partial{\vc{b}}/\partial \Phi) + \Delta \vc{b}'$, which
corresponds to a perturbed polarization sweep of $\psi = \psi_0(\Phi +
\Delta \Phi) + \Delta \psi'$, where $\psi_0(\Phi)$ is the standard
sweep, $\Delta \psi'$ is calculated from $\Delta \vc{b}'$ as
before, and $\Phi \equiv \Omega t$.

As the star rotates, the zeroth-order binormal (as evaluated in
magnetic coordinates) changes both because of the apparent motion of
the observer across the sky and because of the change in the rotating
coordinates themselves.  The change in the binormal is
\begin{equation}
  \frac{d\uvc{b}}{dt} = \frac{db^i}{dt} \bvc{i} + b^i \frac{d\bvc{i}}{dt}
    = \frac{d\uvc{b}^{(1)}}{dt} + \frac{d\uvc{b}^{(2)}}{dt} 
\end{equation}

The first term depends on the motion of the emission point in magnetic
coordinates.  In these coordinates, the zeroth-order binormal only
depends on the azimuth of the emission point, $\phi_0 = \phi_{obs}$.
From (\ref{eq:tan_ph_obs}), this changes in time as
\begin{equation}
  \frac{d\phi_0}{dt} = - \Omega (\cos \alpha + \sin \alpha \cos
  \phi_0 \cot \theta_{obs}).
\end{equation}

Solving for $\theta_{obs}$ in terms of $\theta_0$ gives
\begin{equation}
  \cot \theta_{obs} = \frac{2 - \tan^2 \theta_0}{3 \tan \theta_0}
    \approx \frac{2}{3} \frac{1 - \sin^2 \theta_0}{\sin \theta_0}
    + O(\sin^3 \theta)
\end{equation}

The change in binormal as seen in the rotating frame is then
\begin{eqnarray}
  \frac{d\uvc{b}^{(1)}}{dt} & = & \frac{\partial \uvc{b}}{\partial \phi_0}
      \frac{d\phi_0}{dt} =
      [-\sin \theta_0,\, -\cos \theta_0,\, 0] \frac{d\phi_0}{dt} \\
    & = &
      \Omega \left[
        \begin{array}{c}
          \frac{2}{3} \sin \alpha \cos \phi_0 \left(1 - \sin^2
            \theta_0 \right)+ \cos \alpha \sin \theta_0 \\
          \frac{2}{3} \frac{\sin \alpha \cos \phi_0}{\sin \theta_0}
            \left(1 - \frac{3}{2} \sin^2 \theta_0\right) +
            \cos \alpha \left(1 - \frac{3}{2} \sin^2 \theta_0\right) \\
           0
        \end{array}
      \right] + O(\sin^3 \theta_0)
\end{eqnarray}
where we have expanded up to third order in $\sin \theta_0$.

The second term depends on the rotation of the coordinate system.
Defining $R_y(a), R_z(b)$ to be respectively rotations around the
$y,z$ axes by angles $a,b$, this becomes
\begin{equation}
  \frac{d\uvc{b}^{(2)}}{dt} = R_y(\alpha) R_z(\Omega t)
    \frac{dR_z(-\Omega t)}{dt} R_y(-\alpha) \uvc{b}
\end{equation}
or
\begin{equation}
  \frac{d\uvc{b}^{(2)}}{dt} = \Omega \left[ -\sin \alpha \cos \theta  \cos \phi - 
    \cos \alpha \sin \theta,\, -\cos \alpha \cos \theta +
    \sin \alpha \sin \theta \cos \phi,\, 0 \right]
\end{equation}

Adding these two and expanding to third order in $\sin \theta$ gives
\begin{equation}
  \frac{d\uvc{b}}{dt} = \Omega \left[ -\frac{1}{3} \sin \alpha \cos \phi
    \left(1 - \frac{1}{2} \sin^2 \theta \right),\,
    \frac{2}{3} \frac{\sin \alpha \cos \phi}{\sin \theta},\,
    0 \right] + O(\sin^3 \theta)
\end{equation}

The shift in $\uvc{b}$ due to aberration, (\ref{eq:b1aberr}), then
corresponds to a phase shift of
\begin{equation}
  \Delta \Phi = - 3 \frac{r}{R_L}
\end{equation}
Once this phase shift has been subtracted out, we are left with
a remainder
\begin{equation}
  \Delta \vc{b}' = \frac{r}{R_L} \left[
    \begin{array}{c}
      \frac{5}{3} \cos \alpha \sin \theta -
        \frac{47}{24} \sin \alpha \cos \phi \sin^2 \theta \\
      -\frac{10}{3} \cos \alpha \left(1 -
        \frac{33}{40} \sin^2 \theta\right) +
        \frac{47}{12} \sin \alpha \cos \phi \sin \theta \\
      0
    \end{array} \right] + O(\sin^3 \theta)
\end{equation}
which corresponds to a polarization shift of
\begin{equation}
  \Delta \psi' = - \frac{10}{3} \frac{r}{R_L} \cos \alpha \left(1 -
    \frac{7}{10} \sin^2 \theta \right) +
    \frac{47}{12} \frac{r}{R_L} \sin \alpha \cos \phi \sin \theta
    + O(\sin^3 \theta)
\end{equation}

\end{document}